# Quantum dynamics of a driven two-level molecule with variable dephasing

Samuele Grandi,[1] Kyle D. Major,[1] Claudio Polisseni,[1] Sebastien Boissier,[1] Alex S. Clark,[1, *] and E. A. Hinds[1, †]

[1]*Centre for Cold Matter, Blackett Laboratory, Imperial College London, Prince Consort Road, SW7 2AZ London, UK*

(Dated: July 7, 2016)

The longitudinal ($\Gamma_1$) and transverse ($\Gamma_2$) decay rates of a two-level quantum system have profound influence on its evolution. Atomic systems with $\Gamma_2 = \frac{1}{2}\Gamma_1$ have been studied extensively, but with the rise of solid-state quantum devices it is also important to consider the effect of stronger transverse relaxation due to interactions with the solid environment. Here we study the quantum dynamics of a single organic dye molecule driven by a laser. We measure the variation of $\Gamma_2$ with temperature and determine the activation energy for thermal dephasing of the optical dipole. Then we measure the second-order correlation function $g^{(2)}(\tau)$ of the light emitted by the molecule for various ratios $\Gamma_2/\Gamma_1$ and saturation parameters $S$. We show that the general solution to the optical Bloch equations accurately describes the observed quantum dynamics over a wide range of these parameters, and we discuss the limitations of the various approximate expressions for $g^{(2)}(\tau)$ that appear in the literature.

The two-level atom driven by light is a paradigm for much of quantum physics, producing key quantum phenomena, such as Rabi oscillation between the levels, antibunching of the emitted light, and entanglement between the atom and the light. The system is damped by the loss of photons to the environment and by dephasing of the transition dipole through off-resonant coupling to environmental fluctuations. Population damping can assist in preparing a quantum state, which is important, for example, in quantum memories [1] and quantum gates [2, 3], while dephasing influences quantum effects such as photon-photon interference [4] and can lead to new quantum correlations in solid-state cavity QED [5]. Damping clearly has an important role to play in the development of quantum technologies.

The focus of this letter is on the quantum dynamics of a driven two-level system and the effect that damping has on the intensity correlations of the scattered photons. Early studies of two-level quantum systems used two-level atoms [6] or ions [7] isolated in high vacuum. The upper-state population decays at a rate $1/T_1 = \Gamma_1$, while the transition dipole in these systems decays at half the rate: $1/T_2 = \Gamma_2 = \frac{1}{2}\Gamma_1$. Here $T_1$ is the upper-state lifetime and $T_2$ the coherence time of the system. Such an atom/ion prepared in the excited state subsequently emits a photon of spectral width $\Gamma_1$, and these photons are indistinguishable – a desirable feature for quantum information processing. In seeking to make simpler, more practical, single photon sources there has been a great effort to develop solid-state emitters including quantum dots [8], colour centres in diamond [9] or silicon carbide [10], and our system of choice, organic molecules [11].

Unlike isolated atomic systems, solid-state emitters are coupled to the phonon bath of the solid, which dephases the optical dipole and gives a very large decay rate $\Gamma_2$ at room temperature. Even at 4 K, self-assembled quantum dots and defect centres, being intrinsically bonded to the surrounding crystal, are dephased by lattice phonons [12, 13] that are not fully suppressed. Cavities have been used with quantum dots to increase $\Gamma_1$ to compensate for this dephasing [14–17]. Defect centres in bulk diamond have approached the lifetime-limited linewidth at a temperature of 1.8 K [9].

In contrast, organic molecules can be hosted as impurities in a molecular crystal held together by van der Waals forces. These emitters are often somewhat protected from the lattice phonons and dephase mainly through a local libration of the molecule itself [18, 19]. Consequently, a number of them have exhibited fully suppressed dephasing [20–22]. In the case of dibenzoterrylene (DBT) molecules embedded in an anthracene crystal the dephasing can be frozen out at temperatures as high as 4 K [23]. These molecules then produce a high yield of indistinguishable photons in a 30–40 MHz-wide line at ∼785 nm [24, 25] and could be used to deliver photons very efficiently into a nearby waveguide [26, 27]. In recent experiments we have been able to grow DBT-doped anthracene crystals with an adjustable DBT concentration [28] and to make very thin doped crystals [29], suitable for coupling the molecules to integrated optical structures. We find that some of these molecules can be excited over a trillion times without photo-bleaching [29].

In this work we cool the doped crystal in a cryostat and image a single DBT molecule to investigate its optical properties. We scan the frequency of a pump laser and detect the fluorescence to determine the width of the scattering resonance. By repeating this over a range of sample temperatures we measure the temperature-dependent "transverse" damping rate $\Gamma_2(T)$, and are able to quantify the thermodynamic behaviour of the phonon bath responsible for dephasing the optical dipole. Next we measure $g^{(2)}(\tau)$ – the intensity correlation function of the scattered photons – over a wide range of pump intensities and dephasing rates. In order to interpret our data we solve the optical Bloch equations to find $g^{(2)}(\tau)$ and compare this with our measurements. We show that this molecule does indeed operate as an ideal two-level quantum system when it is driven by resonant light. Although $g^{(2)}(\tau)$ has been measured for many types of quantum emitter, we are not aware of any previous study to inves-tigate the system dynamics as a function of $\Gamma_2$. Various formulae can be found in the literature, each with some range of validity depending on the particular experiment under discussion. Here we show how those formulae relate to the general expression and we specify the approximations that must be valid in each case. We test these using our molecule.

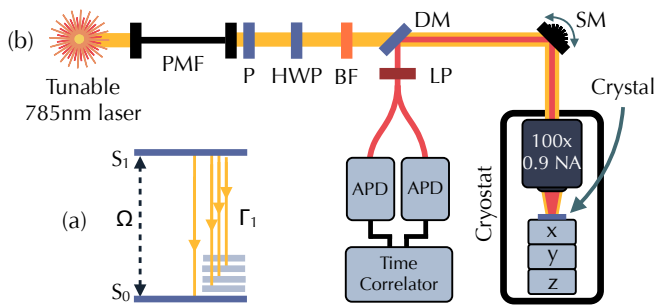

Figure 1. (Color Online) (a) Relevant energy levels for a DBT molecule in anthracene. A laser drives the zero-phonon line, which is also the main decay branch. Some decays go to vibrationally excited levels of the ground electronic state. (b) Schematic diagram of the apparatus. A confocal microscope images fluorescence from a single DBT molecule in the anthracene crystal. PMF: polarisation-maintaining fibre; P: linear polariser; HWP: half-wave plate; BF: band-pass filter; DM: dichroic mirror; SM: steering mirrors; APD: silicon avalanche photodiode; LP: long-pass filter.

The DBT molecule is prepared in a single crystal of anthracene, as described in [28], where the energy levels are as illlustrated in Fig. 1(a). The first electronic excitation, $S_0 \to S_1$, is driven with Rabi frequency $\Omega$ by resonant laser light at 783.73 nm. The spontaneous decay goes directly to the ground state $\sim 40\%$ of the time (the zero-phonon line or ZPL), with the remainder going to vibrationally exited states that relax very rapidly to the ground state.

The apparatus is shown schematically in Fig. 1(b). The crystal is placed on a silicon substrate, which sits on the moveable cold platform of a cryostat (Montana Cryostation) and is thermally connected to it by silver paint. We use integrated heaters to set a sample temperature between 4.0 K and 10.6 K. A room-temperature microscope objective sits 300 μm above the crystal, which it views through a small hole in a thermal radiation shield.

The excitation light is provided by a distributed feedback diode laser, actively locked to a stable, tunable reference cavity. The light is spatially filtered by a single-mode fibre (PMF in Fig. 1), then collimated, polarised and spectrally filtered to remove the background of laser spontaneous emission. A microscope objective lens focusses the beam to a spot of width (FWHM) 550 nm. Two steering mirrors centre this light on the molecule, while its polarisation is aligned with the molecular optical dipole using a half-wave plate. The molecular fluorescence is collected by the same objective and returns along the same path until a dichroic mirror separates the red-shifted vibrational sidebands from the ZPL and back-scattered excitation light. These are further suppressed by a long-pass filter cutting off at 792 nm – between the ZPL and the first sideband. The light is then split into two beams which are detected by silicon avalanche photodiode detectors. The total fluorescence is recorded by summing the two detector signals. A time-correlated counting card, started by one detector and stopped by the other, records the distribution of counting times, which is proportional to $g^{(2)}(\tau)$ (Supplementary Material).

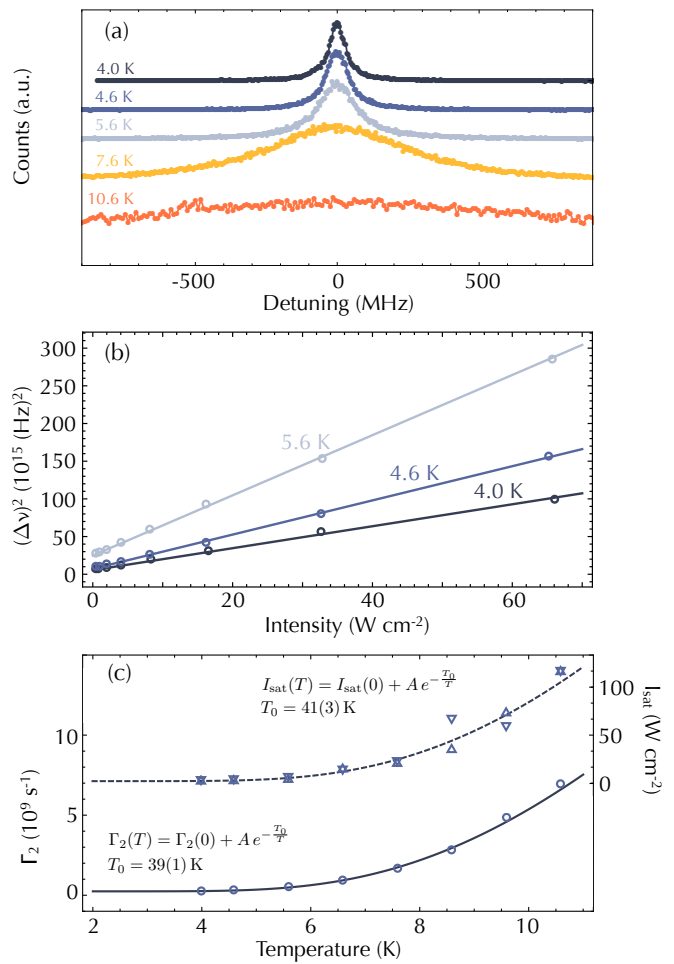

Figure 2. (Color Online) (a) Weakly excited scattering resonances of DBT in anthracene, illustrating increase of line width with temperature. (b) Linewidth-squared versus laser intensity for the 3 lowest temperatures. The intercept gives $\Gamma_2$, while the slope gives $I_{\text{sat}}$. (c) Measured temperature-dependence of $\Gamma_2$ (circles, left ordinate) and $I_{\text{sat}}$ (triangles, right ordinate). Upright triangles are from measurements of $\Gamma_2$ while inverted triangles are from measurements of the peak fluorescence rate. Symbol sizes represent error bars. Zero temperature is uncertain by $\pm 0.25$ K, our uncertainty in temperature difference between the sensor and the molecule.

First, we measure the mean fluorescence rate, proportional to the excited state probability $\rho_{22}$:

$$\rho_{22} = \frac{\frac{1}{2} S}{\delta^2/\Gamma_2^2 + 1 + S}, \qquad (1)$$

where $\delta$ is the angular frequency detuning of the excitation light from resonance and $S = \Omega^2/(\Gamma_1\Gamma_2)$ is the saturation parameter (see Eq.(S.2)). It is convenient to write $S = I/I_{\text{sat}}$, but we cannot measure $I$ at the site of the molecule. Instead we monitor the power $P$ incident on the surface of the anthracene and take $I = 2P/(\pi w^2)$, where $w = 467$ nm is the Gaussian width parameter. Figure 2(a) shows the fluorescence rate varying as we tune $\delta$ through the resonance, and shows that the line has a larger width at higher temperature. According to Eq.(1), the width (in Hz) is $\Delta\nu = (\Gamma_2/\pi)\sqrt{1+S}$. We measure this width at each temperature as a function of laser intensity and extrapolate to zero intensity, as shown in Fig. 2(b) for the three lowest temperatures. The intercepts yield a set of values for $\Gamma_2(T)$, while the slopes give $I_{\text{sat}}(T)$. A separate measure of $I_{\text{sat}}(T)$ is given by the peak scattering rates, which are proportional to $S/(1+S)$. The two methods of determining $I_{\text{sat}}$ give virtually identical results.

The circles in Fig. 2(c) show our results for $\Gamma_2(T)$. These data are well described by the simple temperature dependence $\Gamma_2(T) = \Gamma_2(0) + Ae^{-T_0/T}$, corresponding to thermal activation of a single local phonon mode at energy $k_B T_0$. The line shows the best fit, for which $T_0 = 39(1)$ K. The triangular data points show our values of $I_{\text{sat}}(T)$. When these are described by the same model, $I_{\text{sat}}(T) = I_{\text{sat}}(0) + Be^{-T_0/T}$, for which we plot the best fit, we find $T_0 = 41(3)$ K. Since $I_{\text{sat}} \propto \Gamma_1\Gamma_2$, this common value of $T_0$ indicates that $\Gamma_1$ does not vary significantly over this temperature range, and that the increase in $I_{\text{sat}}$ is just due to the increase of $\Gamma_2$. Our value for $T_0$ is consistent with the lower bound of 35 K reported in [23]. We looked for a spectral feature at that energy (1.7 nm to the red of the ZPL) in the dispersed fluorescence spectrum, but could not see anything above the background.

Next, we investigate the intensity correlation $g^{(2)}(\tau)$ over a wide range of the parameters $S$ and $\Gamma_2$. The left column of Fig. 3 shows three $g^{(2)}$ curves taken at $\Gamma_2 = 2\pi \times 40.5(1)$ MHz, with $S$ adjusted by changing the laser intensity. Panel (a) represents small $S$ where there is a simple dip, (b) shows $S \simeq 1$, where the dip is narrower and $g^{(2)}$ lifts slightly above unity, and (c) shows $S \gg 1$, where there is a clear Rabi oscillation. The right column of Fig. 3 shows curves at $S \simeq 5$ for a range of $\Gamma_2$. In panel (d), $\Gamma_2$ is increased but the Rabi oscillations are still visible. In (e), a small overshoot remains in the wings of the dip, then in (f) we see a simple dip for very large $\Gamma_2$.

All this complexity is contained in the general solution for the two-level system, which we derive in the Supplementary Material from the optical Bloch equations:

$$g^{(2)}(\tau) = 1 - \frac{p+q}{2q}e^{-\frac{1}{2}(p-q)\tau} + \frac{p-q}{2q}e^{-\frac{1}{2}(p+q)\tau}, \quad (2)$$

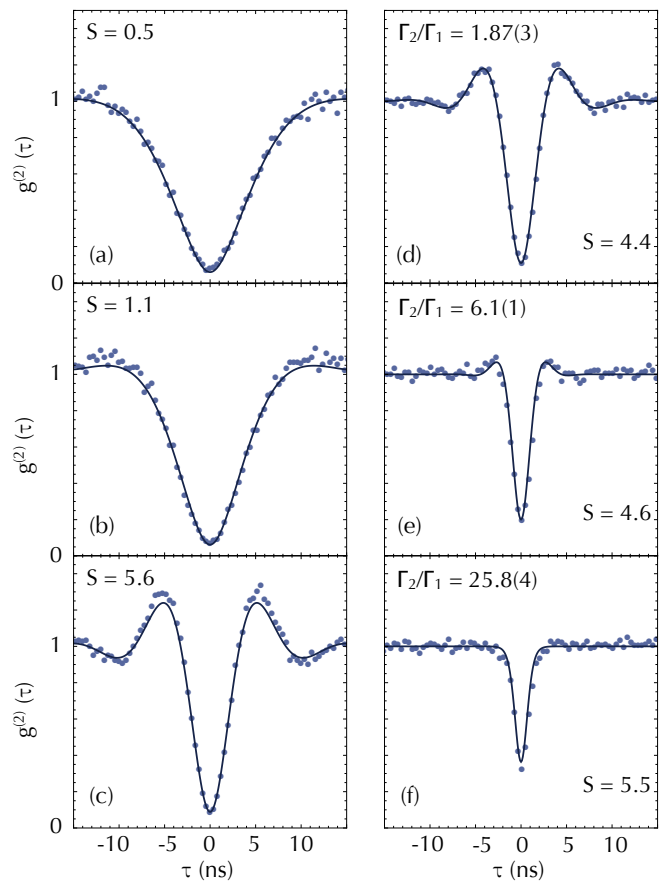

where

$$\begin{aligned} p &= \Gamma_1 + \Gamma_2, \\ q &= \sqrt{(\Gamma_1 - \Gamma_2)^2 - 4\Omega^2}. \end{aligned} \quad (3)$$

Figure 3. (Color Online) Exploration of $g^{(2)}(\tau)$ over a range of the parameters $S$ and $\Gamma_2/\Gamma_1$. (a)–(c) $S$ is varied, with fixed $\Gamma_2/\Gamma_1 = 0.94(2)$. (d)–(f) $\Gamma_2/\Gamma_1$ is varied with $S \simeq 5$. Circles: measured $g^{(2)}$ values. The spread in points is the Poissonian counting noise. Lines: in (b) the line shows the least squares fit to Eq.(2) with $T_1 = 3.70(6)$ ns as the only fitting parameter. In the other panels the line shows Eq.(2) plotted without any free parameters.

To compare our data with this theory, we convolve Eq.(2) with the measured temporal response function of our apparatus, a Gaussian with 455 ps standard deviation. Since all the parameters are known except for $\Gamma_1$, we make a least-squares fit to the data in Fig. 3(b), which yields $T_1 = 3.70(6)$ ns. This is consistent with other values in the literature: $3.3-5.7$ ns [30] and $3.1-3.5$ ns [28]. There are then no more free parameters so the lines in the other plots represent Eq.(2) without any fitting. The agreement is excellent throughout. We are not aware of any other experiment that tests the ability of the optical Bloch equations to describe a real system that is resonantly pumped over a wide range of both $S$ and $\Gamma_2$.

Many experiments in the literature have investigated $g^{(2)}(\tau)$ over a restricted range of parameters, and those papers cite a variety of expressions for $g^{(2)}(\tau)$ that differ from Eq. (2). In the absence of dephasing, $\Gamma_2 = \Gamma_1/2$, and Eq.(2) becomes

$$\lim_{\Gamma_2 \to \frac{1}{2}\Gamma_1} \left[g^{(2)}(\tau)\right] = \\ 1 - e^{-\frac{3}{4}\Gamma_1 \tau} \left(\cosh(\Theta\tau) + \frac{3\Gamma_1}{4\Theta}\sinh(\Theta\tau)\right), \quad (4)$$

where $\Theta = \sqrt{\Gamma_1^2/16 - \Omega^2}$. This is the formula derived by Carmichael and Walls [31]. The low power limit of Eq.(4),

$$\lim_{\substack{\Gamma_2 \to \frac{1}{2}\Gamma_1 \\ \Omega \ll \Gamma_1}} \left[g^{(2)}(\tau)\right] = \left(1 - e^{-\frac{1}{2}\Gamma_1\tau}\right)^2, \quad (5)$$

is given by Loudon [32]. In the presence dephasing, Flagg et al. [33] propose the expression

$$g^{(2)}(\tau)_{\text{Flagg}} = \\ 1 - e^{-\frac{1}{2}(\Gamma_1 + \Gamma_2)\tau}\left(\cos(\mu\tau) + \frac{\Gamma_1 + \Gamma_2}{2\mu}\sin(\mu\tau)\right), \quad (6)$$

where $\mu = \sqrt{\Omega^2 + (\Gamma_1 - \Gamma_2)^2}$. This is not a limit of Eq.(2), but the two do coincide when $\Gamma_2 = \Gamma_1$. A similar formula is proposed by Batalov et al. [34]:

$$g^{(2)}(\tau)_{\text{Batalov}} = \\ 1 - e^{-\frac{1}{2}(\Gamma_1 + \Gamma_2)\tau}\left(\cos(\Omega\tau) + \frac{\Gamma_1 + \Gamma_2}{2\Omega}\sin(\Omega\tau)\right). \quad (7)$$

This too agrees with Eq.(2) at $\Gamma_2 = \Gamma_1$, but is again not a limit of it. In the strong dephasing limit, where $\Gamma_1/\Gamma_2 \ll 1$ and $\Omega/\Gamma_2 \ll 1$, expansion in these small quantities gives $p + q \simeq 2\Gamma_2$, and $p - q \simeq 2\Gamma_1 + 2\Omega^2/\Gamma_2$. With the further approximation that $\Gamma_2\tau \gg 1$, i.e. that the coherence time is much shorter than the measurement time, the third term in Eq.(2) damps away to leave

$$\lim_{\Gamma_2 \gg \{\Gamma_1, \Omega, \tau^{-1}\}} \left[g^{(2)}(\tau)\right] = 1 - e^{-\Gamma_1(1+S)\tau}, \quad (8)$$

showing the single exponential rise common in room temperature measurements of $g^{(2)}(\tau)$ [28, 30].

Figure 4 illustrates the $\{\Gamma_2/\Gamma_1, S\}$ parameter space and plots the regions where Eqs.(4–8) provide good approximations to Eq.(2) (see Supplementary Material). There is a slender region on the left where Eq.(4) is reliable. Equation (5) works over an even smaller region shown inset. Eq.(6) works adequately when close to the condition $\Gamma_2 = \Gamma_1$, while Eq.(7) has a slightly wider range of validity, also centred on this condition. The region of validity for Eq.(8) is on the right of Fig. 4. Over much of the parameter space, Eqs.(4–8) all fail and the full expression of Eq.(2) must be used. Also shown in Fig. 4 are the points corresponding to the measurements of $g^{(2)}(\tau)$ in Fig. 3. We see that measurements (a-c) lie in the region where Eqs.(6,7) are good approximations, while (d) is described by Eq.(7). Measurements (e) and (f) are not well described by any of the approximations but correspond closely to Eq.(2), as shown in Fig.3.

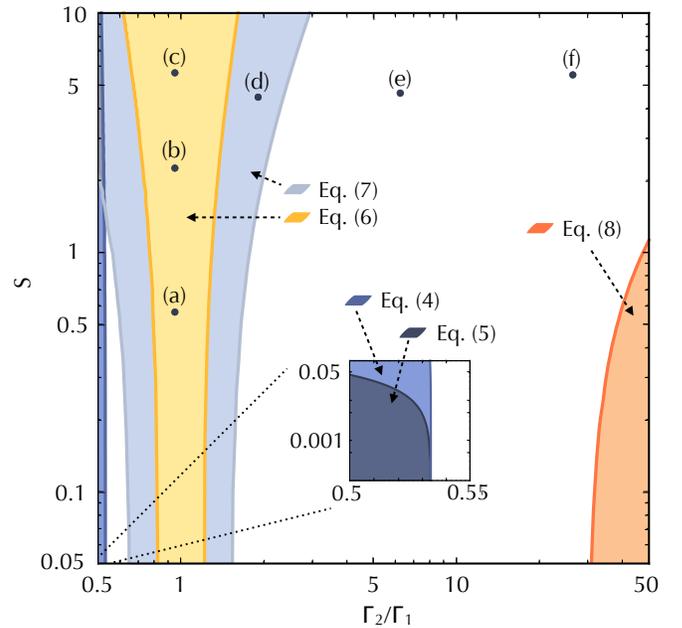

Figure 4. (Color Online) Space of parameters $(\Gamma_2/\Gamma_1, S)$ that determine the behaviour of $g^{(2)}(\tau)$. Shaded areas show parameter ranges where each of the equations (4–8) gives a reliable description of the system dynamics. Points (a)–(f), correspond to the measurements in Fig. 3, which explore the parameter region and establish that Eq. (2) fully describes the system.

In summary, we have investigated the quantum dynamics of a resonantly-driven DBT molecule over a wide range of the parameters $\Gamma_2/\Gamma_1$ and $S$. By measuring how the dephasing rate increases with temperature, we have determined the characteristic energy gap for excitation of the relevant local phonon mode, and we have seen that this mode is not significantly excited in the fluorescence spectrum. We have mapped the evolution of the internal state dynamics by recording $g^{(2)}(\tau)$ and have shown that this is very well described over the broad parameter range by Eq.(2). We also make contact with a variety of other formulae in the literature and show how they relate to the general expression. We conclude that the molecule operates as an ideal two-level quantum system across a wide parameter space. This establishes its suitability for use in photonic circuits, as proposed in [26].

This work was supported in the UK by EPSRC, dstl and the Royal Society, and by the European Commission's Marie-Curie Action-Initial Training Networks: Frontiers in Quantum Technology Project and a Marie Skłodowska Curie Individual Fellowship (Q-MoPS).

* alex.clark@imperial.ac.uk
† ed.hinds@imperial.ac.uk

[1] H. P. Specht, C. Nölleke, A. Reiserer, M. Uphoff, E. Figueroa, S. Ritter, and G. Rempe, Nature **473**, 190 (2011).
[2] A. Reiserer, N. Kalb, G. Rempe, and S. Ritter, Nature **508**, 237 (2014).
[3] T. G. Tiecke, J. D. Thompson, N. P. de Leon, L. R. Liu, V. Vuletić, and M. D. Lukin, Nature **508**, 241 (2014).
[4] J. Iles-Smith, D. P. S. McCutcheon, J. Mørk, and A. Nazir, arXiv:1606.06305 (2016).
[5] J. Iles-Smith and A. Nazir, Optica **3**, 207 (2016).
[6] H. J. Kimble, M. Dagenais, and L. Mandel, Phys. Rev. Lett. **39**, 691 (1977).
[7] F. Diedrich and H. Walther, Phys. Rev. Lett. **58**, 203 (1987).
[8] A. J. Shields, Nature Photonics **1**, 215 (2007).
[9] P. Tamarat, T. Gaebel, J. R. Rabeau, M. Khan, A. D. Greentree, H. Wilson, L. C. L. Hollenberg, S. Prawer, P. Hemmer, F. Jelezko, and J. Wrachtrup, Phys. Rev. Lett. **97**, 083002 (2006).
[10] S. Castelletto, B. C. Johnson, V. Ivády, N. Stavrias, T. Umeda, A. Gali, and T. Ohshima, Nature Materials **13**, 151 (2014).
[11] W. E. Moerner, New Journal of Physics **6**, 88 (2004).
[12] L. Besombes, K. Kheng, L. Marsal, and H. Mariette, Phys. Rev. B **63**, 155307 (2001).
[13] T. Müller, I. Aharonovich, Z. Wang, X. Yuan, S. Castelletto, S. Prawer, and M. Atatüre, Phys. Rev. B **86**, 195210 (2012).
[14] C. Santori, D. Fattal, J. Vučković, G. S. Solomon, and Y. Yamamoto, Nature **419**, 594 (2002).
[15] E. B. Flagg, A. Muller, S. V. Polyakov, A. Ling, A. Migdall, and G. S. Solomon, Phys. Rev. Lett. **104**, 137401 (2010).
[16] R. B. Patel, A. J. Bennett, I. Farrer, C. A. Nicoll, D. A. Ritchie, and A. J. Shields, Nature Photonics **4**, 632 (2010).
[17] N. Somaschi, V. Giesz, L. De Santis, J. C. Loredo, M. P. Almeida, G. Hornecker, S. L. Portalupi, T. Grange, C. Antón, J. Demory, C. Gómez, I. Sagnes, N. D. Lanzillotti-Kimura, A. Lemaítre, A. Auffeves, A. G. White, L. Lanco, and P. Senellart, Nature Photonics **10**, 340 (2016).
[18] P. de Bree and D. A. Wiersma, The Journal of Chemical Physics **70**, 790 (1979).
[19] W. H. Hesselink and D. A. Wiersma, The Journal of Chemical Physics **73**, 648 (1980).
[20] W. P. Ambrose, T. Basché, and W. E. Moerner, The Journal of Chemical Physics **95**, 7150 (1991).
[21] S. Kummer, T. Basché, and C. Bräuchle, Chemical Physics Letters **232**, 414 (1995).
[22] F. Jelezko, B. Lounis, and M. Orrit, The Journal of Chemical Physics **107**, 1692 (1997).
[23] A. A. L. Nicolet, P. Bordat, C. Hofmann, M. A. Kol'chenko, B. Kozankiewicz, R. Brown, and M. Orrit, ChemPhysChem **8**, 1929 (2007).
[24] A. A. L. Nicolet, P. Bordat, C. Hofmann, M. A. Kol'chenko, B. Kozankiewicz, R. Brown, and M. Orrit, ChemPhysChem **8**, 1215 (2007).
[25] J.-B. Trebbia, H. Ruf, P. Tamarat, and B. Lounis, Opt. Express **17**, 23986 (2009).
[26] J. Hwang and E. A. Hinds, New Journal of Physics **13**, 85009 (2011).
[27] N. R. Verhart, G. Lepert, A. L. Billing, J. Hwang, and E. A. Hinds, Opt. Express **22**, 19633 (2014).
[28] K. D. Major, Y.-H. Lien, C. Polisseni, S. Grandi, K. W. Kho, A. S. Clark, J. Hwang, and E. A. Hinds, Review of Scientific Instruments **86**, 083106 (2015).
[29] C. Polisseni, K. D. Major, S. Boissier, S. Grandi, A. S. Clark, and E. A. Hinds, Opt. Express **24**, 5615 (2016).
[30] C. Toninelli, K. Early, J. Bremi, A. Renn, S. Götzinger, and V. Sandoghdar, Opt. Express **18**, 6577 (2010).
[31] H. J. Carmichael and D. F. Walls, Journal of Physics B: Atomic and Molecular Physics **9**, 1199 (1976). Eq. 6.5.
[32] R. Loudon, *The Quantum Theory of Light*, 3rd ed. (Oxford Science Publications, 2000) p. 350. Eq. 8.3.9.
[33] E. B. Flagg, A. Muller, J. W. Robertson, S. Founta, D. G. Deppe, M. Xiao, W. Ma, G. J. Salamo, and C. K. Shih, Nature Physics **5**, 203 (2009). Eq. 2.
[34] A. Batalov, C. Zierl, T. Gaebel, P. Neumann, I. Y. Chan, G. Balasubramanian, P. R. Hemmer, F. Jelezko, and J. Wrachtrup, Physical Review Letters **100**, 077401 (2008). Eq 1.

# Supplementary Material: Quantum dynamics of a driven two-level molecule with variable dephasing

Samuele Grandi,[1] Kyle D. Major,[1] Claudio Polisseni,[1] Sebastien Boissier,[1] Alex S. Clark,[1, *] and E. A. Hinds[1, †]

[1]*Centre for Cold Matter, Blackett Laboratory, Imperial College London, Prince Consort Road, SW7 2AZ London, UK*

(Dated: July 7, 2016)

## DERIVATION OF EQ.(2) FOR $g^{(2)}(\tau)$

The evolution of a damped 2-level system, driven at resonance by a coherent field $E_0 \cos(\omega t)$, can be described by the optical Bloch equations

$$\dot{\rho}_{22}(t) = i\frac{\Omega}{2}[\rho_{12}(t) - \rho_{21}(t)] - \Gamma_1 \rho_{22}(t),$$
$$\dot{\rho}_{21}(t) = i\frac{\Omega}{2}[\rho_{11}(t) - \rho_{22}(t)] - (\Gamma_2 - i\delta)\rho_{21}(t),$$
$$\dot{\rho}_{12}(t) = \dot{\rho}_{21}(t)^*,$$
$$\dot{\rho}_{11}(t) = -\dot{\rho}_{22}(t). \tag{S.1}$$

Here $\rho_{ij}(t)$ are the density matrix elements, with $\rho_{11}$ being the population of the stable lower level, while the upper level population $\rho_{22}$ has spontaneous decay rate $\Gamma_1$, and the total population $\rho_{11} + \rho_{22}$ is conserved. The strength of the excitation is characterised by $\hbar\Omega = dE_0$, where $d$ is the electric dipole transition matrix element. The off-diagonal density matrix elements are damped at rate $\Gamma_2$, and $\delta$ is the detuning of the optical frequency from the resonant frequency. The time-dependence of the interaction has been eliminated by making the rotating wave approximation and transforming to an appropriate interaction picture.

The steady-state solution of these equations gives the excited-state population as

$$\rho_{22}(\infty) = \frac{\frac{1}{2}\left(\frac{\Omega^2}{\Gamma_1 \Gamma_2}\right)}{\frac{\delta^2}{\Gamma_2^2} + 1 + \left(\frac{\Omega^2}{\Gamma_1 \Gamma_2}\right)}. \tag{S.2}$$

Since $\Omega^2$ is proportional to the intensity $I$ of the light that drives the excitation, one normally defines a "saturation intensity" $I_{sat}$ such that

$$\frac{\Omega^2}{\Gamma_1 \Gamma_2} = \frac{I}{I_{sat}} = S. \tag{S.3}$$

This ratio $S$ is known as the saturation parameter. When a laser is scanned across the scattering resonance, the scattering rate is proportional to $\rho_{22}$, and hence the peak of the resonace signal is proportional $\frac{S}{1+S}$, while the width of the resonance (FWHM) is $\Gamma_2\sqrt{1+S}$.

We turn now to the time-dependence of the density matrix, which we evaluate here for the special case of resonant excitation, $\delta = 0$. Writing $\rho_{12}$ as $\frac{1}{2}(u - iv)$, Eqs.(S.1) simplify to

$$\dot{\rho}_{22}(t) = \frac{\Omega}{2}v(t) - \Gamma_1 \rho_{22}(t),$$
$$\dot{v}(t) = \Omega(\rho_{11}(t) - \rho_{22}(t)) - \Gamma_2 v(t),$$
$$\dot{\rho}_{11}(t) = -\dot{\rho}_{22}(t). \tag{S.4}$$

The real part of $\rho_{12}$, i.e. $u$, separates from these equations because the light is resonant, and therefore the driven dipole is in quadrature with the driving field. We are interested in solving these equations to find $\rho_{22}(t)$, with the initial condition that $\rho_{22}(0) = 0$. This solution is found by taking the Laplace transform of Eq.(S.4):

$$s\tilde{\rho}_{22}(s) = \frac{\Omega}{2}\tilde{v}(s) - \Gamma_1\tilde{\rho}_{22}(s),$$
$$s\tilde{v}(s) = \Omega(\tilde{\rho}_{11}(s) - \tilde{\rho}_{22}(s)) - \Gamma_2\tilde{v}(s),$$
$$s\tilde{\rho}_{11}(s) - 1 = -s\tilde{\rho}_{22}(s). \tag{S.5}$$

On solving these equations, we obtain

$$\tilde{\rho}_{22}(s) = \frac{\frac{1}{2}\Omega^2}{2s(s^2 + \Gamma_1\Gamma_2 + s(\Gamma_1 + \Gamma_2) + \Omega^2)} \tag{S.6}$$

With the substitutions

$$p = \Gamma_1 + \Gamma_2,$$
$$q = \sqrt{(\Gamma_1 - \Gamma_2)^2 - 4\Omega^2}, \tag{S.7}$$

we re-write Eq.(S.6) as

$$\tilde{\rho}_{22}(s) = \frac{c_1}{s - s_1} + \frac{c_2}{s - s_2} + \frac{c_3}{s - s_3}, \tag{S.8}$$

where

$$\{c1, c2, c3\} = \{\frac{2\Omega^2}{p^2 - q^2}, -\frac{\Omega^2}{q(p-q)}, \frac{\Omega^2}{q(p+q)}\},$$
$$\{s1, s2, s3\} = \{0, -\frac{1}{2}(p - q), -\frac{1}{2}(p + q)\}. \tag{S.9}$$

The inverse transform then gives the desired solution,

$$\rho_{22}(t) = c_1 + c_2 e^{s_2 t} + c_3 e^{s_3 t}. \tag{S.10}$$

At long times, the second and third terms damp away, leaving the steady-state solution

$$\rho_{22}(\infty) = c_1 = \frac{\frac{1}{2}\Omega^2}{\Gamma_1\Gamma_2 + \Omega^2}. \tag{S.11}$$

This is, of course, the $\delta = 0$ case of Eq.(S.2).

The second-order correlation function of the radiation field is defined as [1]

$$g^{(2)}(\tau) = \frac{\left\langle :\hat{I}(t)\hat{I}(t+\tau): \right\rangle}{\left\langle :\hat{I}(\tau): \right\rangle^2}, \quad (S.12)$$

where $\hat{I}(t)$ is the intensity operator, and the whole function is expressed in normal ordering. This is the normalised joint probability of detecting a photon at time $t+\tau$, given that another photon was detected at time $t$. Because the probability of detecting a photon is directly proportional to the probability $\rho_{22}$ of the system being excited, and because the first detection projects the system into the lower level, Eq.(S.10) provides exactly what we need to evaluate $g^{(2)}(\tau)$:

$$g^{(2)}(\tau) = \frac{\rho_{22}(\tau)}{\rho_{22}(\infty)} = 1 - \frac{p+q}{2q}e^{-\frac{1}{2}(p-q)\tau} + \frac{p-q}{2q}e^{-\frac{1}{2}(p+q)\tau} \quad (S.13)$$

This is the result we quote in Eq.(2) in the main text.

## COMPARISON WITH OTHER $g^{(2)}$ FORMULAE

To compare Eq.(S.13) with the various approximate formulae in the literature we consider the normalised mean-square difference

$$\mathcal{D}_a(S, \Gamma_2/\Gamma_1) = \frac{\int_{-\infty}^{\infty} \left(g^{(2)}(\tau) - g_a^{(2)}(\tau)\right)^2 d\tau}{\int_{-\infty}^{\infty} \left(1 - g^{(2)}(\tau)\right)^2 d\tau}, \quad (S.14)$$

where $g^{(2)}(\tau)$ is the function given in Eq.(S.13), and $g_a^{(2)}(\tau)$ is the formula under consideration. If $\mathcal{D}_a > 10^{-3}$, we find typically that the two functions disagree noticeably when plotted. We therefore take the area of agreement to be the region that satisfies $\mathcal{D}_a < 10^{-3}$, and those are the areas that we plot in Fig. 4 of the main text.

---

[1] R. Loudon, *The Quantum Theory of Light*, 3rd ed. (Oxford Science Publications, 2000) p. 176.